\documentclass[sigconf,nonacm]{acmart}

\AtBeginDocument{%
  }

\setcopyright{acmlicensed}
\copyrightyear{2018}
\acmYear{2018}
\acmDOI{XXXXXXX.XXXXXXX}
\acmConference[Conference acronym 'XX]{Make sure to enter the correct
  conference title from your rights confirmation email}{June 03--05,
  2018}{Woodstock, NY}
\acmISBN{978-1-4503-XXXX-X/2018/06}

\usepackage{multirow}

\usepackage{bbm}

\begin{document}

\title[Comparing GNN-Based and ID-Based Item Embeddings in the Yandex Ecosystem]{Embedding Items at Scale: Comparing GNN-Based and ID-Based Item Embeddings in the Yandex Ecosystem}

\author{Sergei Makeev}
\email{neuralsrg@gmail.com}
\orcid{0009-0003-5451-6475}
\affiliation{%
  \institution{Yandex}
  \city{Moscow}
  \country{Russia}
}

\author{Artem Matveev}
\email{matfu21@ya.ru}
\orcid{0009-0004-0271-221X}
\affiliation{%
  \institution{Yandex}
  \city{Moscow}
  \country{Russia}
}

\author{Vladimir Baikalov}
\email{nonameuntitled159@gmail.com}
\orcid{0009-0009-4864-2305}
\affiliation{%
  \institution{Yandex}
  \city{Moscow}
  \country{Russia}
}

\author{Kirill Khrylchenko}
\email{elightelol@gmail.com}
\orcid{0009-0007-3640-8795}
\affiliation{%
  \institution{Yandex}
  \city{Moscow}
  \country{Russia}
}

\renewcommand{\shortauthors}{Makeev et al.}

\begin{abstract}
  Transformer-based sequential recommendation models, which process sequences of user-item interactions, rely heavily on the item embedding strategy. Existing approaches either use pretrained item embeddings or learn them end-to-end with the transformer. To the best of our knowledge, no prior work has compared these options from both cost and quality perspectives in a large-scale industrial setting. This paper is a case study that compares pretrained industrial graph neural network item embeddings with end-to-end trainable item embeddings across two mature production recommendation systems at Yandex: Yandex Market and Yandex Music. We additionally evaluate both approaches on a low-resource dataset sampled from Yandex Lavka production logs, for which both the data and code are publicly available for demonstration purposes. Our results show that a separate pretraining stage helps when training data is limited, but provides no worthwhile benefit for large-scale models trained on extensive datasets.
\end{abstract}

\ccsdesc[500]{Information systems~Personalization}
\ccsdesc[500]{Information systems~Learning to rank}
\ccsdesc[500]{Information systems~Recommender systems}

\keywords{Deep Learning, Personalization, Sequential Recommendation, Graph Representation Learning}


\maketitle

\section{Introduction}
Sequential recommendation aims to predict the next item a user is likely to interact with based on their past interactions. Transformer-based models have been widely adopted for this task in large-scale industrial systems such as Pinterest\,\cite{pancha2022pinnerformer, xia2023transact, xia2025transact, chen2025pinfm}, ByteDance\,\cite{chai2025longer, zhang2025onetrans}, Taobao\,\cite{chen2019behavior}, Kuaishou\,\cite{liu2024kuaiformer}, Zalando\,\cite{celikik2022reusable}, Yandex\,\cite{khrylchenko2023personalized}, and others. These models take as input a chronologically ordered sequence of user-item interactions and use it to produce personalized recommendations.

One way to encode items in these sequences is to map their IDs to the embeddings learned jointly with the transformer without a separate embedding training stage. However, modern recommender systems operate on large and highly dynamic item catalogs, making it impractical to assign a unique embedding to every item ID. Instead, the hashing trick\,\cite{weinberger2009feature} maps each item ID to an index in a fixed-size embedding table using a hash function. To reduce collisions, the multihash technique\,\cite{tito2017hash} applies multiple hash functions to the same item ID, retrieves several table entries, and combines them by summation\,\cite{tito2017hash, agarwal2024omnisearchsage} or concatenation\,\cite{coleman2023unified}. \citet{coleman2023unified} report deploying such multihash embeddings in various web-scale search, ads, and recommendation models at Google. 

Another way to encode items is to use pretrained item embeddings. User interactions can be represented as a heterogeneous graph where nodes are users and items, and edges denote relationships such as clicks, purchases, and likes. Graph representation learning aims to learn high-quality node embeddings that can later be used as item embeddings in the transformer model. \citet{pancha2022pinnerformer, xia2023transact, xia2025transact} report deploying such multi-stage systems at Pinterest. Following \citet{hamilton2017representation}, graph representation learning can be viewed as an encoder-decoder problem, where the encoder maps nodes to embeddings and the decoder reconstructs the graph structure. Graph neural networks (GNNs) can be divided into two groups:

\begin{enumerate}
    \item \textbf{Transductive GNNs} directly train node embeddings and require all nodes to be present during training. As new nodes appear, such models must be retrained. Albeit simple, TwHIN is an efficient model of this type, proposed by \citet{el2022twhin} and deployed at X (formerly Twitter). The encoder maps the set of nodes \(\mathcal{V}\) and relations \(\mathcal{R}\) to trained embeddings: \(\text{ENC}(\cdot) = \theta_{\cdot} \in \mathbb{R}^d\), where \(d\) is the embedding dimension. Thus, the trained parameters can be viewed as an embedding matrix \(\Theta \in \mathbb{R}^{(|\mathcal{V}|+|\mathcal{R}|)\times d}\). The decoder reconstructs the probability logits of an edge \(e = (s, r, t)\), formed by two nodes \(s, t \in \mathcal{V}\) and a relation \(r \in \mathcal{R}\), being present in the graph: \(\text{DEC}(e) = \text{DEC}(\text{ENC}(s), \text{ENC}(r), \text{ENC}(t)) = (\theta_s + \theta_r)^T\theta_t\). The model is trained to maximize the log-likelihood of predicting a binary "real" or "fake" label for observed edges in a heterogeneous graph \(\mathcal{G}\) and negatively sampled edges \(\mathcal{N}(\cdot)\):
    \[\arg\max_\Theta \sum_{e \in \mathcal{G}}\left[\log \sigma(\text{DEC}(e)) + \sum_{e^{\prime} \in \mathcal{N}(e)}\log \sigma(-\text{DEC}(e^{\prime}))\right],\] where \(\mathcal{N}(s, r, t) = \{(s, r, t^{\prime}): t^{\prime} \in \mathcal{V}\} \cup \{(s^{\prime}, r, t): s^{\prime} \in \mathcal{V}\}\) is the set of negative edges obtained by replacing either the source or the target node of the positive edge.
     \item \textbf{Inductive GNNs} learn a generalizable, parameterized function rather than embeddings for specific nodes. This function aggregates information from a node local neighbors based on their features. As a result, the full graph is not required during training, and previously unseen nodes can be handled naturally. MultiBiSage\,\cite{gurukar2022multibisage} extends PinSage\,\cite{ying2018graph}, which was used for item embeddings in Pinterest production models\,\cite{pancha2022pinnerformer, xia2023transact, xia2025transact}, and adapts it to heterogeneous graphs. A heterogeneous graph \(\mathcal{G}\) can be decomposed into multiple bipartite graphs \(\mathcal{G}_r\), each containing a single relation type \(r \in \mathcal{R}\). In MultiBiSage, the encoder processes visual and textual features of a node \(s \in \mathcal{V}\) and its neighbors in each bipartite graph \(\mathcal{G}_r\) using a transformer to obtain intermediate representations. These representations are then aggregated across all bipartite graphs by another transformer to form a final embedding \(e_s^{\theta}\), where \(\theta\) denotes encoder parameters. The decoder learns the probability distribution \(p(t \mid s)\), \(t, s \in \mathcal{V}\), of node \(t\) being "similar" to \(s\) over the item catalog using a sampled softmax:
    \[\arg\max_\theta \sum_{(s, t)}\log \left( \frac{\exp\{ \langle e_s^{\theta}, e_t^{\theta}\rangle - \log Q(t)\}}{\sum_{t^{\prime} \in \mathcal{N}}\exp\{ \langle e_s^{\theta}, e_{t^{\prime}}^{\theta} \rangle - \log Q(t^{\prime})\}} \right),\]  where \(\mathcal{N}\) is a set of in-batch and uniformly sampled negatives \(t^{\prime}\), known as Mixed Negative Sampling\,\cite{yang2020mixed}, and \(\log Q(t)\) is the logQ correction term\,\cite{yi2019sampling}.
\end{enumerate}

This paper presents a case study comparing pretrained TwHIN and MultiBiSage item embeddings (referred to as \textbf{GNN embeddings}) with multihash item ID embeddings trained end-to-end with the transformer from scratch (referred to as \textbf{ID embeddings}) across two mature recommendation systems at Yandex, as well as on a public dataset, to extend the conclusions to low-resource scenarios. We analyze both cost and quality, and assess whether the additional pretraining stage is worthwhile.
\section{Method}
Sequential recommendation aims to recommend the most relevant item to a user based on their interaction history \(x^u=\left(i^u_{1}, i^u_{2}, ..., i^u_{l_u}\right)\), where \(l_u\) is the number of interactions for user \(u\).

In our experiments, we focus on ranking models. Following the production ranking setup proposed by \citet{khrylchenko2023personalized}, we split training of the transformer ranker into pretraining and fine-tuning stages. In both stages, the user is represented by their interaction history \(x^u\).

\textit{Model architecture.} We use a production two-tower architecture, where the user tower is a transformer over the interaction sequence \(x^u\), and the item tower is a residual network.

\textbf{The item tower} combines the item title with the item embedding (GNN or ID embedding). An embedding bag layer maps the BPE-tokenized title into token embeddings and sums them. The BPE vocabulary contains \(\mathcal{O}(10^5)\) tokens. The item embedding is added to the title representation. A linear layer then projects the resulting embedding into a higher-dimensional space, followed by three residual blocks. Each block consists of a linear layer, ReLU activation, dropout, and Layer Normalization. A final linear layer maps the embedding back to the original dimension, and the output is \(L_2\)-normalized.

\textbf{The user tower} encodes the user interaction history. First, the item tower is applied to each item in the interaction sequence. Learnable action embeddings (the sets of possible actions are platform-dependent) and positional embeddings are added element-wise, and a CLS token is concatenated to the sequence. A bidirectional transformer encoder is then applied, and the \(L_2\)-normalized output corresponding to the CLS token is used as the user representation.

A dot product between user and item embeddings, equivalent to cosine similarity, is used as the relevance score.

\textit{Pretraining} is formulated as a next-item prediction task. For each \(k\), given a subsequence \(\left(i^u_{1}, i^u_{2}, ..., i^u_{k-1}\right)\), the model predicts \(i^u_{k}\). A sampled softmax loss with in-batch negatives is used to learn a probability distribution over the item catalog:
\[
\mathcal{L}_{pretrain}(u, i^u_{k}, \mathcal{N}) = - \log \frac{\exp\{r_{ui^u_{k}} / \tau\}}{\exp\{r_{ui^u_{k}} / \tau\} + \sum_{n \in \mathcal{N}} \exp\{r_{un} / \tau\}},
\]
where \(r_{ui}\) is the relevance score between user \(u\) and item \(i\), \(\mathcal{N}\) is a set of in-batch negatives, and \(\tau\) is a temperature parameter.

\textit{Fine-tuning} trains the model to rank candidate items. For each user request, the recommender system returns a set of \(s^u\) items \(\mathcal{I}^u= \{i^u_{1}, ..., i^u_{s^u}\}\) shown to user \(u\). For each of these items, user actions \(\mathcal{A}^u= \{a^u_{1}, ..., a^u_{s^u}\}\) are stored and used to calculate both pointwise and pairwise ranking losses.

Pointwise ranking is formulated as a click prediction task:
\[
\mathcal{L}_{pt}(u, i) = -t_{ui} \log f_{ui}^{pt} - (1 - t_{ui}) \log(1 - f_{ui}^{pt}),
\]
where \(f_{ui}^{pt} = \sigma(\alpha\cdot r_{ui} + \beta)\), \(\alpha, \beta\) are trained parameters, \(r_{ui}\) is the relevance score, and \(t_{ui} = \mathbbm{1}\left\{ a^u_{i} = \text{"click"} \right\}\).

For pairwise ranking, we construct all pairs \((i^u_p, i^u_n)\) between items \(i^u_p\) we treat as positive interactions and items \(i^u_n\) with which the user did not interact. Following \citet{bai2023regression}, the pairwise loss is:
\[
\mathcal{L}_{pr}(u, i^u_p, i^u_n) = -\log \frac{f_{up}^{pr}}{f_{up}^{pr} + f_{un}^{pr}},
\]
where \(f_{ui}^{pr} = \sigma(\delta\cdot r_{ui} + \gamma)\), and \(\delta, \gamma\) are trained parameters.

The fine-tuning loss is a weighted sum of pointwise and pairwise losses:
\[
\mathcal{L}_{finetune}(u, \mathcal{I}^u, \mathcal{A}^u) = \sum_{i \in \mathcal{I}^u} \mathcal{L}_{pt}(u, i) + \frac{1}{10} \sum_{(i^u_p, i^u_n)} \mathcal{L}_{pr}(u, i^u_p, i^u_n).
\]

To compare GNN and ID embeddings, we plug each embedding type into the item tower and train the transformer model. The fine-tuned model outputs ranking scores, which are then used, together with other statistical features, as input to a CatBoost\,\cite{prokhorenkova2018catboost} production ranker. To evaluate the contribution of the transformer model, we train the CatBoost ranker twice: once with these ranking scores and once without them, and report the relative difference in metrics. In all experiments, we use a timestamp-based strategy to split the training and test data.

To assess statistical significance, we split the CatBoost dataset into 32 equal folds. For each fold, we train both the baseline CatBoost and the CatBoost with the additional feature on that fold and compute metrics on the remaining data. We compare the results using the Wilcoxon signed-rank test and report differences that are significant at \(p < 0.01\).
\section{Experiments}
We compare pretrained GNN item embeddings, trained before the transformer ranker, with embeddings trained end-to-end from scratch (ID embeddings) on three Yandex platforms: Yandex Market (e-commerce), Yandex Music (music streaming), and Yandex Lavka (grocery e-commerce). For Yandex Market and Yandex Music, we experiment with production models described in\,\cite{khrylchenko2023personalized, khrylchenko2025scaling} and large-scale in-house datasets to draw conclusions for large-scale models. For Yandex Lavka, we train a smaller model on a low-resource dataset; both the data and code are available on our GitHub (see \autoref{sec:lavka}). When comparing GNN and ID embeddings, we obtain them for the same set of items.

\subsection{Yandex Market}

We formulate the following research questions:
\begin{itemize}
    \item \textbf{RQ1:} Do pretrained GNN embeddings lead to better ranking quality than ID embeddings in a large-scale domain?
    \item \textbf{RQ2:} Does combining both embedding approaches provide worthwhile improvements?
\end{itemize}

\textit{Data.} We collect user logs over a one-year period from Yandex Market, which serves millions of users. The data contains \(\mathcal{O}(10^7)\) items, and user actions include clicks, cart additions, likes, and purchases. Test data is collected over two days following the training period.

We use TwHIN and MultiBiSage to train GNN item embeddings. Embeddings are computed for the 17\% most popular items, which cover most user interactions; all other items share a single embedding.

\textit{TwHIN.} We construct a heterogeneous user-item graph with edges representing clicks, orders, likes, and cart additions.

\textit{MultiBiSage.} Here, we construct two bipartite graphs: an item-cart graph (connecting items to the carts they were added to) and an item-order graph (connecting items to completed orders). To reduce popularity bias, we prune the graphs by limiting the number of edges per node to at most several thousand. We also exclude cart and order nodes whose normalized entropy exceeds a threshold. Entropy is computed over item categories, and normalization is performed with respect to cart or order size, rather than the number of categories. For each bipartite graph, we use a 4-layer transformer over visual and textual representations. We then aggregate vectors from the two graphs with a 2-layer transformer. Following \citet{gurukar2022multibisage}, we sample 50 neighbors per node using random walks to define the local neighborhood. We then construct positive item pairs, where the source item is the item viewed by the user and the target item is the item clicked from the similar-items recommendations.

\textit{ID embeddings.} We use an embedding matrix with \(\mathcal{O}(10^6)\) entries and 64 dimensions. For the multihash technique, we use 6 lookups. The corresponding embeddings are concatenated and projected to the initial dimension.

The transformer processes sequences of up to 256 events. To address RQ2, we combine TwHIN and ID embeddings in the item tower and study their joint effect. We report offline relative nDCG differences for three recommendation surfaces: retargeting, discovery, and recommendations from the cart page. Retargeting is an unconstrained personalization setting, similar to eBay’s Recently Viewed Items module\,\cite{wang2021personalized}, while discovery restricts recommendations to previously unseen items.

The results in \autoref{tab:market} show that ID embeddings trained end-to-end from scratch outperform pretrained GNN embeddings in this large-scale setting. Combining GNN and ID embeddings yields additional gains, but we do not consider the extra GNN training cost to be worthwhile.

\begin{table}
    \caption{Relative nDCG difference on large-scale Yandex Market data and production model. Results are statistically significant.}
    \label{tab:market}
    \begin{tabular}{l|ccc}
        \toprule
        Item embeddings & Discovery & Cart & Retargeting \\
        \midrule
        No (only content info) & +0.506\% & +0.103\% & +0.565\% \\
        \midrule
        TwHIN & +0.790\% & +0.151\% & +0.943\% \\
        MultiBiSage & +0.565\% & +0.122\% & +0.651\% \\
        ID embeddings & \underline{+1.238\%} & \underline{+0.215\%} & \underline{+1.486\%} \\
        \midrule
        TwHIN + ID embeddings & \textbf{+1.273\%} & \textbf{+0.235\%} & \textbf{+1.522\%} \\
        \bottomrule
    \end{tabular}
    \Description{Table compares TwHIN, MultiBiSage, and end-to-end trained item ID embeddings in the production model on the large-scale Yandex Market dataset.}
\end{table}

\subsubsection{Compute Time and Memory Analysis}
All experiments were conducted on 8 NVIDIA A100 GPUs rented via a third-party compute provider. For ID embeddings, training time depends on embedding dimensionality, number of lookups, and embedding table size. It ranges from 40 hours (16 dimensions, 2 lookups) to 70 hours (256 dimensions, 6 lookups), and from 41 hours for \(\mathcal{O}(10^4)\) embedding table entries to 52 hours for \(\mathcal{O}(10^6)\) entries in our experimental setup (64 dimensions, 6 lookups). MultiBiSage training takes 52 hours, while TwHIN training takes 5 hours.

During the main model training, we store frozen GNN embeddings on SSD using Lightning Memory-Mapped Database (LMDB), while learnable ID embeddings are kept on GPU. Due to additional RAM-GPU communication overhead, we do not report training time changes from using GNN embeddings.

\subsection{Yandex Music}

We formulate the following research questions:
\begin{itemize}
    \item \textbf{RQ3:} Do the comparison results between the two embedding approaches generalize to another large-scale domain?
    \item \textbf{RQ4:} Does fine-tuning pretrained GNN item embeddings improve performance?
\end{itemize}

\textit{Data.} The dataset contains hundreds of billions of interactions from \(\mathcal{O}(10^7)\) users and over \(\mathcal{O}(10^6)\) tracks. User actions include listening time, likes, and skips.

We train embeddings for the same \(\mathcal{O}(10^5)\) most popular tracks, which account for more than 90\% of user interactions. Since textual and visual content is less informative for tracks than in e-commerce, and given the weak MultiBiSage results in the previous experiment, we only train TwHIN item embeddings.

\textit{TwHIN.} In the graph, nodes are users and tracks, and edges correspond to likes and long plays (at least 90\% of the track played).

\textit{ID embeddings.} Since we only embed \(\mathcal{O}(10^5)\) items, we train unique embeddings for all of them without hashing.

We train the production transformer ranker\,\cite{khrylchenko2025scaling} on sequences of up to 2048 interactions. We report the relative differences in pair accuracy and weighted pair accuracy, where the latter reweights actions based on the importance of each action type. To answer RQ4, instead of combining both embedding strategies, we fine-tune pretrained TwHIN item embeddings end-to-end with the transformer model.

As shown in \autoref{tab:music}, ID embeddings significantly outperform TwHIN embeddings. Interestingly, even after fine-tuning, TwHIN embeddings do not improve the final performance.

\begin{table}
    \caption{Relative accuracy difference on large-scale Yandex Music data and production model. Results are statistically significant.}
    \label{tab:music}
    \begin{tabular}{l|ccc}
        \toprule
        \begin{tabular}[l]{@{}l@{}}Item \\ embeddings \end{tabular} & \begin{tabular}[l]{@{}l@{}}Embeddings \\ are fine-tuned \end{tabular} & \begin{tabular}[l]{@{}l@{}}Pair \\ Accuracy \end{tabular} & \begin{tabular}[l]{@{}l@{}}Weighted Pair \\ Accuracy \end{tabular} \\
        \midrule
        TwHIN & - & +0.348\% & +0.325\% \\
        TwHIN & \checkmark & \underline{+0.524\%} & \underline{+0.448\%} \\
        ID embeddings & \checkmark & \textbf{+0.699\%} & \textbf{+0.603\%} \\
        \bottomrule
    \end{tabular}
    \Description{Table compares TwHIN and end-to-end trained item ID embeddings in the production model on the large-scale Yandex Music dataset.}
\end{table}

\subsection{Yandex Lavka}
\label{sec:lavka}

For this experiment, we release data and code on our GitHub\footnote{https://github.com/matfu-pixel/Comparing-GNN-based-and-ID-based-Item-Embeddings-Web-Scale-Perspective}. We investigate the following research question:
\begin{itemize}
    \item \textbf{RQ5:} Does training GNN item embeddings yield improvements in a limited-resource setting?
\end{itemize}

\textit{Data.} The data, sampled over a one-year period, comes from the Yandex Lavka grocery e-commerce service. The dataset contains 15 million user-item interactions from 3315 users and 25833 items. Available actions are views, clicks, cart additions, and purchases.

To keep the published dataset size manageable, we do not include textual or image descriptions of items and restrict our study to TwHIN GNN embeddings. Due to the relatively small number of unique items, we train both \textit{TwHIN} and \textit{ID embeddings} for all items, without popularity-based clipping. We train TwHIN on a user-item graph, where edges correspond to clicks, cart additions, and purchases.

The ranking transformer processes sequences of up to 256 events. We report nDCG@5, nDCG@10, and nDCG@20 per user request. Unlike the previous experiments, transformer scores are used directly to rank items and are not passed as additional CatBoost features.

\begin{table}
    \caption{Experiment results on low-resource Yandex Lavka data and small transformer model. Results are averaged over 10 runs.}
    \label{tab:lavka}
    \begin{tabular}{l|cccc}
        \toprule
        \begin{tabular}[l]{@{}l@{}}Item \\ embeddings \end{tabular} &
        \begin{tabular}[l]{@{}l@{}}Embeddings \\ are fine-tuned \end{tabular} &
        \begin{tabular}[l]{@{}c@{}}nDCG \\ @5 \end{tabular} &
        \begin{tabular}[l]{@{}c@{}}nDCG \\ @10 \end{tabular} &
        \begin{tabular}[l]{@{}c@{}}nDCG \\ @20 \end{tabular} \\
        \midrule
        TwHIN & - & \underline{0.337} & \underline{0.409} & \underline{0.457} \\
        TwHIN & \checkmark & \textbf{0.342} & \textbf{0.415} & \textbf{0.464} \\
        ID embeddings & \checkmark & 0.333 & 0.406 & 0.456 \\
        \bottomrule
    \end{tabular}
    \Description{Table compares TwHIN and end-to-end trained item ID embeddings in a small-scale model on the open low-resource Yandex Lavka dataset.}
\end{table}

In contrast to the previous experiments, the results in \autoref{tab:lavka} show that models using pretrained TwHIN embeddings consistently outperform those using only ID embeddings. This indicates that when training data is limited, an additional GNN training stage improves model performance.
\section{Conclusion}

We compared two item embedding approaches for ranking: pretrained graph neural network embeddings and embeddings trained end-to-end with the transformer from scratch. We ran experiments on three Yandex platforms: Yandex Market, Yandex Music, and Yandex Lavka. For Yandex Market and Yandex Music, we used production models and large-scale training datasets. For Yandex Lavka, we used a smaller dataset and a compact model, and released both the dataset and code. Our results show that in low-resource settings, pretraining item embeddings is beneficial, while in large-scale settings pretrained embeddings do not provide benefits that justify their additional cost.

\bibliographystyle{ACM-Reference-Format}
\bibliography{sample-base}


\end{document}